# Breast Cancer Neoadjuvant Chemotherapy Treatment Response Prediction Using Aligned Longitudinal MRI and Clinical Data

Rahul Ravi[1], Ruizhe Li[1], Tarek Abdelfatah[2], Stephen Chan[2], Xin Chen[1]

[1] School of Computer Science, University of Nottingham, Nottingham, UK

[2] Nottingham City Hospital, Nottingham, UK

**Abstract**

**Aim**: This study investigates treatment response prediction to neoadjuvant chemotherapy (NACT) in breast cancer patients, using longitudinal contrast-enhanced magnetic resonance images (CE-MRI) and clinical data. The goal is to develop machine learning (ML) models to predict pathologic complete response (PCR binary classification) and 5-year relapse-free survival status (RFS binary classification).

**Method**: The proposed framework includes tumour segmentation, image registration, feature extraction, and predictive modelling. Using the image registration method, MRI image features can be extracted and compared from the original tumour site at different time points, therefore monitoring the intratumor changes during NACT process. Four feature extractors, including one radiomics and three deep learning-based (MedicalNet, Segformer3D, SAM-Med3D) were implemented and compared. In combination with three feature selection methods and four ML models, predictive models are built and compared.

**Results**: The proposed image registration-based feature extraction consistently improves the predictive models. In the PCR and RFS classification tasks logistic regression model trained on radiomic features performed the best with an AUC of 0.88 and classification accuracy of 0.85 for PCR classification, and AUC of 0.78 and classification accuracy of 0.72 for RFS classification.

**Conclusions:** It is evidenced that the image registration method has significantly improved performance in longitudinal feature learning in predicting PCR and RFS. The radiomics feature extractor is more effective than the pre-trained deep learning feature extractors, with higher performance and better interpretability.

## 1 Introduction

Breast cancer is the most common cancer in women in 157 out of 185 countries [1]. Neoadjuvant chemotherapy treatment (NACT), which involves giving chemotherapy before surgery, is regarded as a potential "standard of care approach", as it reduces tumour bulk and enables breast conservation. Complete tumour resolution at surgery, known as pathological complete response (PCR), has a high likelihood of achieving cure [2]. However, not all patients achieve PCR owing to a range of factors, affecting the growth and characteristics of the tumour. Additionally, of those who achieve PCR, recurrence is likely. Relapse free survival (RFS) is also a crucial clinical outcome in patient management. Accurately predicting PCR and RFS for individual patients provides significant clinical value for personalised treatment planning and improving treatment outcomes. Contrast-enhanced Magnetic Resonance Imaging (CE-MRI) becomes a standard clinical imaging modality for tumour localisation and quantification. CE-MRIs are normally acquired prior to NACT (known as baseline) and after NACT is completed (endpoint). Many existing studies have shown the effectiveness of extracting features from the baseline CE-MRI for PCR prediction. Very few studies investigated and utilised the changes of imaging features between the baseline and end-point MRI for PCR and RFS prediction. The changes of imaging features provide additional information for more accurate clinical outcomes predictions and better treatment monitoring. However, longitudinal feature extraction from different time-points of MRI of the same patient poses technical challenges, due to soft tissue deformation, tumour size changes, image acquisition protocol differences, etc.

The key aim of this work is to develop a longitudinal MRI feature extraction framework utilizing advanced image registration methods for PCR and RFS prediction in NACT. This framework includes tumour segmentation, image registration, feature extraction, feature selection and predictive modelling.

## 2 Related Works

There are numerous research works that use machine learning (ML) method for PCR and RFS prediction using different sets of features, including clinical features, radiology reports, radiomics images, and deep learning-based feature representations. Additionally, different imaging modalities were used, including MRI, mammogram, and ultrasound images.

One approach developed by Eben et al. [3], named RESONATE, combines radiomic and deep learning-based models to extract features from both tumoral and peritumoral tumour regions in baseline MRI for PCR classification. By fusing prediction results from two classifiers using logistic regression, the RESONATE approach achieved an area under the curve (AUC) of 0.79, surpassing models that relied on either radiomic or deep learning methods [3]. This integration demonstrates the benefit of leveraging multiple analytic representations from different tumoral regions to improve predictive performance.

In another study, Dell'Aquila et al [4] used demographic and tumour subtype data alongside radiology reports from a cohort of 475 patients for PCR prediction. They implemented several ML models, including logistic regression and random forests, achieved AUCs ranging from 0.74 to 0.76 for PCR prediction and up to 0.85 for overall survival. This research emphasises the importance of incorporating diverse datasets and socio-economic factors in enhancing model performance. It also shows that tumour subtypes and imaging characteristics are critical predictors of PCR and OS. Although model performance is good, authors stated that their sample size was small (240) and 'only MRI radiology reports' were used instead of images.

The development of a deep learning radiomic nomogram (DLRN) was explored by Jiang, M. et al [5], who utilised pre- and post-treatment ultrasound images for predicting PCR. The DLRN model, which combined deep learning and handcrafted radiomic features, achieved an AUC of 0.94 in the validation cohort [5]. This study demonstrated the effectiveness of integrating deep learning with radiomic features and clinicopathological risk factors to enhance predictive accuracy.

Multiparametric MRI's role in predicting PCR was highlighted by Mohamed, R.M. et al [6], who analysed radiomic models based on multiparametric MRI for triple-negative breast cancer (TNBC). Their top-performing model, combining 35 radiomic features of relative difference between mid-point and baseline MRI, achieved an AUC of 0.905 in the training set and 0.802 in the testing set. The high performance of this model supports the use of multiparametric MRI for early prediction of treatment response in TNBC [6]. However, their study only focuses on TNBC, limited generalizability to other sub-types, and a lower AUC during testing could indicate model overfitting.

In a study by Park, J. et al [7], the integration of ML models with features from different MRI sequences was explored. The research utilised tumoral and peritumoral features across various MRI phases, with the K-Nearest Neighbour model achieving the highest AUC of 0.9631. This study demonstrates the advantage of combining features from different MRI phases and regions to improve PCR prediction accuracy. However, it lacks analysis of disease progression as it only uses pre-treatment MRIs. The diagnostic accuracy of using MRI for predicting PCR across different molecular subtypes was analysed by Kim, J. et al [8]. Their study found that the accuracy varied significantly by subtype, with HR−/HER2− showing the highest accuracy and HR−/HER2+ the lowest. The study also highlighted challenges related to subtle residual enhancement, which can lead to false-negative results. This study suggests the need for subtype-specific approaches and careful interpretation of MRI findings.

Lastly, Huang, Y. et al [9] developed ensemble learning models using longitudinal multiparametric MRI, extracting radiomic and deep learning features. Their ensemble model, which integrated pre-, post, and delta-MRI features [9], achieved high AUCs of up to 0.974 and demonstrated strong performance in three external validation cohorts [9]. This research shows the potential of combining multiple feature types and longitudinal MRIs to enhance the prediction accuracy for PCR.

SPAG5 expression is an important biomarker that signifies tumour growth. Overexpression (value of 1) of SPAG5 could indicate a low chance of survival [10]. This is found in studies conducted by Abdel-Fatah, T.M. *et al.* [11] and Canu, V. *et al* [12]. It demonstrates that a low SPAG5 reduces cancer cell proliferation. Hence, this important biomarker is included in our study.

In summary, recent research works have made significant efforts in predicting PCR and survival using various ML and imaging techniques. The integration of diverse data sources, including radiomic and deep learning features across different MRI sequences and timepoints, has proven to be beneficial. However, all existing works extracted image features from the tumour site and the residual tumour site only. For the patients where the tumours are significantly downsized and eventually disappear in MRI, conventional feature extraction techniques that extract features from the residual tumour site can generate unreliable and value of zeros (due to no tumour remaining). However, the texture information at the original tumour site could provide valuable information for PCR and RFS prediction. Locating the original tumour site in the follow-up MRI images requires accurate image alignment (i.e., image registration). None of the existing studies investigated the benefit of extracting features consistently from the original tumour site at different stages of the NACT for PCR and RFS prediction. Note that a disappeared tumour does not necessarily mean pathological complete response, hence it is still useful to use endpoint MRIs to predict PCR.

The contributions of this paper are summarised as follows: (1) it is the first study that utilising advanced image registration techniques to extract image features from longitudinal CE-MRIs for PCR and RFS predictions in NACT; (2) the predictive models were based on a combination of shape, texture, and clinical features on a large local dataset (n=665) compared to other works. (3) Experimental results have shown a significant performance improvement of using the image registration-based longitudinal feature extraction method compared to the method without image registration.

## 3 Methods

This proposed framework contains four main stages: tumour segmentation, image registration, feature extraction, and predictive modelling to predict treatment response in breast cancer patients administered with NACT, as shown in Figure 1. Detailed descriptions are provided in the following subsections.

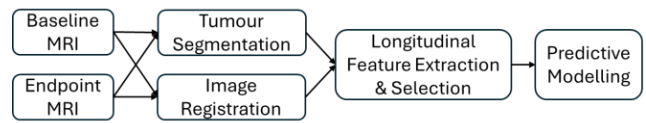

Figure 1: Overview of the proposed framework.

### 3.1 Dataset
A total of 665 patients undertaking NACT at Nottingham University Hospital (UK) were included in this study. A set of clinical measurements were collected prior to treatment, ER, PgR, HER2, TNM staging, and SPAG5 expression. A subset of patients that missed clinical measures, without MRI images or with incomplete RFS follow-ups were excluded from the study. The final included number of subjects and their key statistics are reported in Tables 1 for PCR and RFS.

Table 1: Summary data statistics for PCR and RFS

| PCR | | | |
|---|---|---|---|
| Characteristic | PCR | No-PCR | Total |
| dataset (%) | 167 (27.9%) | 431 (72.1%) | 598 |
| SPAG5+ | 142 (56.6%) | 153 (43.4%) | 295 |
| SPAG5- | 25 (8.3%) | 278 (91.7%) | 303 |
| ER+ | 66 (20.0%) | 263 (80.0%) | 329 |
| ER- | 101 (37.5%) | 168 (62.5%) | 269 |
| HER2+ | 92 (46.7%) | 105 (53.3%) | 197 |
| HER2- | 75 (18.7%) | 326 (81.3%) | 401 |
| RFS | | | |
| Characteristic | RFS | No-RFS | Total |
| dataset (%) | 121 (36.5%) | 227 (63.5%) | 348 |
| SPAG5+ | 54 (35.5%) | 98 (64.5%) | 152 |
| SPAG5- | 67 (34.2%) | 129 (65.8%) | 196 |
| ER+ | 56 (30.6%) | 127 (69.4%) | 183 |
| ER- | 65 (39.4%) | 100 (60.6%) | 165 |
| HER2+ | 19 (17.0%) | 93 (83.0%) | 112 |
| HER2- | 102 (43.2%) | 134 (56.8%) | 236 |

The CE-MRIs of each patient was acquired prior to NACT (baseline: timepoint A) and at the completion cycle of NACT (endpoint: timepoint B). All MRI scans are anonymised and ethically approved to be used in this study. All tumours in all CE-MRIs were segmented using a semi-automatic segmentation software [13]. These tumour masks were validated by a breast oncologist and used in the feature extraction stage.

### 3.2 Longitudinal Image Registration

Image registration is a process that transforms the source image to match the target image using an estimated displacement field [14]. A popular application of image registration is the analysis of longitudinal medical images. The objective of image registration in this study is to analyse the textural changes of the original tumour site between the baseline and endpoint MRI scans. In this study, the source and target images were baseline and endpoint MRI scans, respectively. The displacement field allows the tumour mask of baseline MRI to be warped to the endpoint MRI coordinate space to identify the region of the original tumour site in the endpoint MRI.

We adopt our previously developed unsupervised image registration method [15]. It is comprised of three main components: encoder-decoder structure, displacement field estimation, and a spatial transformer. The encoder-decoder is a multi-level network, which is used to extract feature maps from the input volumes using 3D convolution operators. The encoder and decoder have symmetric structure that using 3D convolutional operators for feature extraction and feature map up-sampling respectively. Additionally, skip connections are applied to connect the encoder and decoder. At each training iteration, a displacement field is estimated to warp the source image to the target coordinate using the spatial transformer layer. The warped source image is then compared to the target image for loss calculation using normalised cross correlation and a smoothness term. The image registration model is optimised by minimising the loss function using a set of paired source-target training images. For more technical details, please refer to the original paper [15].

In our study, all paired MRI scans were resized to 128×128×128 for image registration model training. It is found to be computational efficient without sacrificing registration accuracy. Note that, the output warped source mask is resized to the original target image dimension and aligned to the original target image for feature extraction, hence no information loss occurred in the feature extraction stage. Figure 2 shows a paired source and target MRIs, and the alignment of warped source image and the target image. It can be seen from Figure 2 (c) that the original tumour site highlighted in pink can be identified in the target image.

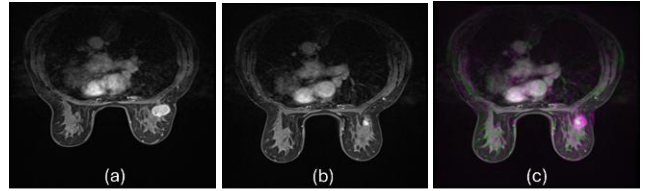

Figure 2: Image registration results. (a) Source image: baseline MRI. (b) Target image: endpoint MRI; (c) warped source image aligned to the target image coordinate. The pink region indicates the original tumour site.

### 3.3 Feature Extraction

Feature extraction is the process by which key information about intratumoral and peritumoral regions is gathered from the MRI scans and corresponding labels. Pre-feature extraction, the dataset consisted of timepoints 'A' (baseline), 'B' (endpoint) and warped 'A' volumes and labels. All labels were used to extract bounding boxes representing the region of interest in the MRIs (i.e. the tumour). This bounding box was used to extract a cube from the respective MRIs. This enabled effective feature extraction through improved representation of the texture features of the tumoral region. These cubes were then resized to 128x128x128 so that the tumour region could be analysed using pre-trained deep learning-based extractors. Two types of feature extraction methods were employed in this study: radiomics (pre-defined features) and deep learning (self-learned features based on a large image dataset). Radiomic features were extracted using the PyRadiomics package [16]. Additionally, three different pretrained deep learning feature extractors were compared, SAM-Med3D [17], MedicalNet [18] and SegFormer3D [19], which were shown to be effective in many clinical applications.

PyRadiomics [16] offers a wide range of features that can be extracted from medical images (2D) and volumes (3D). These features include geometric, first order statistical features (e.g. mean, standard deviation, histogram-based, etc), higher order texture features (e.g. grey-level co-occurrence matrix, grey-level zone matrix, grey-level run length matrix, neighbouring grey tone difference matrix, and grey-level dependence matrix features, etc.). All features are extracted from the MRI volume by specifying a region of interest (ROI) indicated by the tumour mask.

SAM-Med3D is a 3D transformer-based segmentation model designed for general-purpose medical image analysis [17]. Unlike standard Segment Anything Models, which operate on 2D slices, SAM-Med3D processes the entire 3D volumes, capturing spatial dependencies across all three dimensions. It was trained on a large-scale dataset of 143,000 medical images and masks, encompassing various imaging modalities and anatomical regions. The model architecture follows a vision transformer design, consisting of 12 transformer layers with an embedding dimension of 384 and 8 attention heads. During feature extraction, the image encoder outputs a volumetric feature map of shape (384, 8, 8, 8) for each MRI scan. This feature map is processed by global average pooling and then flattened, resulting in a compact feature representation with size of 384×1.

MedicalNet is a 3D convolutional neural network based on the ResNet architecture and developed specifically for transfer learning across diverse medical imaging tasks such as classification and segmentation [18]. It is implemented as a 3D ResNet-200, which consists of 50 layers organised into four stages of residual blocks utilising 3D convolutions. The network follows an encoder-decoder paradigm that enables hierarchical feature extraction across multiple scales. MedicalNet was pretrained on a broad array of 3D medical datasets, covering different organs and modalities like CT and MRI, to promote generalisation across domains. For feature extraction, the encoder generates feature maps of shape (2048, 16, 16, 16), which are then average pooled across spatial dimensions to obtain a reduced tensor of shape (2048, 1, 1, 1). This tensor is then flattened to produce a final feature vector of size 2048×1.

SegFormer3D is a lightweight hierarchical transformer designed specifically for 3D medical image segmentation, balancing computational efficiency and segmentation performance [19]. Unlike conventional vision transformers that rely heavily on global self-attention, SegFormer3D employs a combination of multi-scale attention and MLP-based decoding, making it suitable for use on smaller medical datasets. The architecture comprises four encoder stages with increasing feature dimensionality, using embedding sizes of 32, 64, 160, and 256. It incorporates depth-wise separable convolutions and hierarchical down-sampling to reduce memory usage without sacrificing spatial information. During feature extraction, the encoder outputs a tensor of shape (1, 3, 128, 128, 128), representing the batch, channel, and spatial dimensions. This tensor is average pooled along the channel dimension to obtain a representation of shape (1, 1, 128, 128, 128). Adaptive average pooling is then applied to reduce this tensor to a final shape of (128, 1, 1), which is flattened to produce a feature vector of shape 128×1.

We compared feature extraction with and without image registration of the longitudinal MRIs. "without registration" features were produced using timepoint 'A' and 'B'. For both timepoint A and B, features were extracted from the tumour region defined by a bounding box seperately. However, there were timepoint 'B' cases where the tumours no longer exist (i.e. responded to the NACT). In these cases, a cube of zeros of shape 128x128x128 representing an empty tumour label was used for feature extraction. Subsequently, the aforementioned four feature extractors were applied to 'A' and 'B' separately to produce the "without registration" set of features and then combined using (A-B)/A, measuring the ratio of changes.

For "with registration" feature extraction, image registration provides a solution by transforming the tumour region from timepoint 'A' into the coordinate space of the timepoint 'B'. Therefore, the warped tumour mask represents the location of the original tumour site. This region was then used at time point B for feature extraction. This avoids the situation of no features are extracted when the tumour is absence. To combine features of both timepoints, (A-B)/A is also used.

Note that for geometric feature extraction using the radiomics method, the "without registration" is used as it represents the true tumour geometry at timepoint B. The "with registration" method is only applied for extracting textural features.

Additional to the MRI-based features, four clinical features were identified as important factors in PCR and RFS prediction, including oestrogen receptor status, HER2 amplification status, TNM staging and SPAG5 expression.

### 3.4 Predictive Modelling

Four machine learning models, including random forest (RF), logistic regression (LR), kernel support vector machine (kSVM) and multi-layer perceptron neural network (MLP), were used for predictive modelling of PCR and RFS classifications. Pre-processing stages include feature normalisation, feature selection, and missing values imputation.

### 3.4.1 Feature Selection

The non-binary features were first normalised using min-max normalisation method. During feature selection and dimensionality reduction, feature meaning preservation is important because identification of crucial features can inform decision making. For PCR and RFS event classification, three feature selection methods were used: recursive feature elimination (RFE), random forest ranking, and the chi-squared test. RFE iteratively removes the least important features based on logistic regression coefficients, refining the subset until only the most relevant remain [21]. Random forest ranks features by their contribution to reducing Gini impurity across trees, effectively capturing non-linear patterns and class separation [21]. The chi-squared test assesses the association between each feature and the class label by comparing observed and expected frequencies, with higher scores indicating stronger relevance.

Given the limited training data, to avoid curse of dimensionality and model overfitting, we aim to retain the minimal number of features for predictive modelling. The final retained feature set is presented in the evaluation section.

### 3.4.2 Machine Learning Modelling

Four ML models [23], including RF, LR, kSVM, and MLP were used for PCR and RFS classifications. RF is an ensemble of decision trees (DT) that improves generalisation by averaging predictions across multiple DTs. DT a rule-based model that splits data into a tree structure using feature thresholds. It is known for its interpretability and ability to model non-linear relationships. Four parameters (criterion, max depth, min samples per leaf, and min samples per split) were used to fine-tune the model. LR is a linear model, which is efficient and interpretable. The hyperparameters include regularisation type and strength, solver, and max iterations. kSVM seeks an optimal hyperplane that maximises the margin between classes. Parameters used included kernel type, regularisation weight, polynomial degree, and maximum iterations. MLP is a deep neural network that captures complex, non-linear patterns through multiple hidden layers. The hyperparameters include layer shape, regularisation weight, learning rate, and max iterations. The hyperparameters of these ML models were fine-tuned using a validation set.

## 4 Method Evaluation

There are a total of eight feature sets (from four feature extractors, each extracting "with registration" and "without registration" features). For each feature set, the combinations of three feature selection methods (section 3.4.1) and four ML models (section 3.4.2) were applied and compared. All models were evaluated using 5-fold cross validation and the results presented are the average of 8 runs using different random seeds to split the data. Classification accuracy were used as the metric for hyperparameter tuning. Classification accuracy and AUC-ROC were used for method comparison. Furthermore, the Wilcoxon signed rank test was used to measure the statistical significance between the paired methods of with/without registration. A derived p-value less than 0.05 was used to indicate statistical significance.

### 4.1 Parameter Settings

The four machine learning models were optimised and resulted in the following best hyperparameters. The RF used 100 estimators, Gini impurity, max depth of 20, and both minimum samples per leaf and split set to 4. LR was trained with L2 regularisation weight of 0.4, the lbfgs solver, and 7000 iterations. The kSVM classifier used a regularisation strength of 10, and 15000 iterations. The MLP classifier utilised three hidden layers (each with 100 neurons), ReLU activation, the Adam optimiser, an alpha value of 0.0001, a learning rate of 0.001, and 15,000 iterations.

### 4.2 Evaluation of Longitudinal Image Registration

The image registration network MrRegNet [15] was trained on pairs of timepoint A-B MRI scans for fifty epochs, learning rate of 0.00001, a training batch size of 1 and an evaluation batch size of 1. Both breasts in each MRI were manually segmented to evaluate the registration accuracy. Dice coefficient is used as the evaluation metric to compare the breast region of the target image and the warped breast region of the source image. MrRegNet also accepts the breast mask as a guidance to help with the image registration. The key hyperparameter settings (smoothness term weights at multi-scale) and the corresponding registration performance are listed in Table 2. The model with mask guidance and the smoothness weight of (96, 48, 24, 16) have shown to achieve the best performance, hence was used for the subsequent stages. An example of aligned MRI images is shown in Figure 2.

Table 2: Image registration performance measured by Dice, mean± standard deviation are presented.

| Smoothness Weights | Mask Guided | Dice |
|---|---|---|
| 48,24,16,8 | No | 0.9327 ± 0.020 |
| 96,48,24,8 | No | 0.9354 ± 0.010 |
| **96,48,24,16** | **Yes** | **0.9498 ± 0.020** |
| 128,96,64,32 | Yes | 0.9456 ± 0.025 |
| 128,96,64,32 | Yes | 0.9474 ± 0.030 |

### 4.3 PCR and RFS Prediction Using Baseline MRI

The experimental results of using imaging features extracted from baseline MRI and clinical features for PCR and RFS classification are presented in Table 3 and Table 4 respectively. The "Feature Extractor" column shows the best feature extractor that associated with each of the four ML models.

It can be seen from the results that the radiomic feature extractor combined with LR model achieved the best result for PCR classification (AUC=0.82 and classification accuracy=0.78). For the RFS classification task, the radiomics features combined with LR model achieved the best performance (AUC=0.70, accuracy of 0.68), which is significantly better than other methods (P<0.05).

Overall, the radiomics feature extractor outperforms the deep learning based feature extractors and LR model consistently achieved the best performance for both PCR and RFS predictions. For PCR prediction, the retained features are flatness, GLCM correlation, difference average, difference entropy, Imc2, sum entropy, NGTDM strength, ER status, HER2 amplification status, SPAG5 expression. For RFS prediction, the retained features are first-order entropy, kurtosis, GLCM correlation, maximum probability, sum entropy, GLRLM run entropy, run variance, GLSZM grey level non-uniformity (normalized), zone entropy, ER status, HER2 amplification status, SPAG5 expression, TNM tumour stage.

Table 3: PCR classification using baseline MRI and clinical features. Mean± standard deviation are presented.

| ML Model | Feature Extractor | Accuracy | AUC |
|---|---|---|---|
| RF | Radiomic | 0.77±0.01 | 0.78±0.01 |
| **LR** | **Radiomic** | **0.78±0.01** | **0.82±0.02** |
| SVM | Radiomic | 0.77±0.01 | 0.80±0.01 |
| MLP | Radiomic | 0.76±0.02 | 0.79±0.01 |

Table 4: RFS classification using baseline MRI and clinical features. Mean± standard deviation are presented.

| ML Model | Feature Extractor | Accuracy | AUC |
|---|---|---|---|
| RF | Radiomic | 0.67±0.01 | 0.68±0.02 |
| **LR** | **Radiomic** | **0.68±0.02** | **0.70±0.01** |
| SVM | Radiomic | 0.66±0.01 | 0.68±0.01 |
| MLP | MedicalNet | 0.66±0.01 | 0.69±0.01 |

## 4.4 PCR and RFS Prediction Using Longitudinal MRIs

The experimental results of using imaging features extracted from the longitudinal MRIs (both time points A and B), and clinical features for PCR and RFS classifications are presented in Table 5 and Table 6, respectively.

In Table 5, it can be observed that, by using the longitudinal feature for both with and without image alignment, the models improved the performance compared to the baseline models in Table 3. However, with the help of image alignment (W columns in Table 5), all ML models achieved better results than the without alignment models (statistical significance indicated by the p values). The radiomics features combined with LR model achieved the best performance in PCR with AUC of 0.86 and classification accuracy of 85%.

The retained features of the best PCR model (LR+ radiomics with image alignment) include least axis length, maximum 2D diameter column, maximum 2D diameter row, first-order 10th percentile, 90th percentile, maximum, mean absolute deviation, median, root-mean squared, GLCM correlation, GLDM dependence non-uniformity, small dependence emphasis, GLRLM long run high grey level emphasis, run length non-uniformity, run percentage, GLSZM grey level variance, size zone non-uniformity (normalized), ER status, HER2 amplification status, SPAG5 expression.

Table 5: PCR classification using longitudinal MRIs and clinical features. Mean± standard deviation are presented.

| ML Model | Feature Extractor | Accuracy W | Accuracy WO | AUC W | AUC WO | p |
|---|---|---|---|---|---|---|
| RF | Radiomic | 0.81 ±0.01 | 0.80 ±0.01 | 0.85 ±0.01 | 0.83 ±0.01 | 0.02 |
| **LR** | Radiomic | **0.85 ±0.02** | **0.83 ±0.04** | **0.88 ±0.01** | **0.86 ±0.01** | **0.01** |
| SVM | Radiomic | 0.83 ±0.01 | 0.81 ±0.02 | 0.87 ±0.01 | 0.86 ±0.03 | 0.03 |
| MLP | Radiomic | 0.83 ±0.01 | 0.82 ±0.05 | 0.87 ±0.02 | 0.85 ±0.01 | 0.02 |

Table 6: RFS classification using longitudinal MRIs and clinical features. Mean± standard deviation are presented.

| ML Model | Feature Extractor | Accuracy W | Accuracy WO | AUC W | AUC WO | p |
|---|---|---|---|---|---|---|
| RF | Radiomic | 0.72 ±0.01 | 0.71 ±0.01 | 0.74 ±0.02 | 0.72 ±0.02 | 0.01 |
| **LR** | Radiomic | **0.72 ±0.01** | **0.70 ±0.02** | **0.78 ±0.01** | **0.76 ±0.01** | **0.01** |
| SVM | Radiomic | 0.71 ±0.02 | 0.70 ±0.01 | 0.78 ±0.02 | 0.75 ±0.01 | 0.01 |
| MLP | Medical Net | 0.72 ±0.01 | 0.70 ±0.01 | 0.75 ±0.01 | 0.74 ±0.01 | 0.02 |

For the RFS prediction results in Table 6, it can be seen that the longitudinal feature models with/without image alignment all achieved improved performance compared to the corresponding baseline models in Table 4. With image alignment, the performance is further improved with statistical significance. The LR model using radiomics features achieves the best performance with 13 retained features, including first-order root mean squared, GLCM difference variance, inverse variance, joint entropy, GLRLM run percentage, GLSZM high grey level zone emphasis, size zone non-uniformity (normalized), small area emphasis, small area high grey level emphasis, zone percentage, ER status, HER2 amplification status, TNM tumour stage.

## 5 Discussion and Conclusions

This study investigated different modern image feature extraction methods, and produced machine learning models based on longitudinal feature representations for PCR and RFS prediction in NACT breast cancer. Particularly, the image segmentation and registration methods provide a robust foundation for inspecting tumour region changes and aligning longitudinal MRI volumes. Longitudinal feature extraction produces valuable information about the tumour and surrounding regions, which significantly impacts the modelling performance.

The key findings from this study are three folds. (1) Pre-defined radiomics features achieved better performance than pre-trained deep learning feature extractors. This could be due to the fact that the dimensionality of the deep learning features is much higher than the radiomic features given this relatively small dataset, which requires significant dimensionality reduction. More importantly, radiomics features are more interpretable than the deep learning features for clinical applications. (2) To apply image registration on longitudinal MRIs prior to feature extraction significantly help the feature learning process, resulting in significantly improved prediction power for both PCR and RFS. (3) Longitudinal feature extraction and modelling achieved significantly improved prediction accuracy compared to using baseline image alone, which indicates the importance and benefit of monitoring treatment responses using imaging techniques.

Accurate prediction of PCR and RFS can offer critical insights for clinicians, potentially leading to more personalized and adaptive treatment planning. These findings not only validate the technical pipeline but also demonstrate the transformative potential of AI in breast oncology.